# Searching for seismo-ionospheric earthquakes precursors: Total electron content disturbances before 2005-2006 seismic events

Yu.V. Romanovskaya[1], A.A. Namgaladze[1], O.V. Zolotov[1], N.A. Starikova[2], V.Z. Lopatiy[2]

[1] *Polytechnic Faculty of MSTU, Physics Department*
[2] *Faculty of Physics, Mathematics, Computer Science and Programming, Murmansk State Humanitarian University*

**Abstract.** During earthquakes preparation periods significant disturbances in the ionospheric plasma density are often observed. These anomalies are caused by lithosphere-atmosphere-ionosphere interaction, particularly by the seismic electric field penetrating from the ground surface into the ionosphere. The seismic electric field produces electromagnetic $\boldsymbol{E}\times\boldsymbol{B}$ drift changing plasma density over the epicenter region and magnetically conjugated area. The paper is devoted to analysis of regular Global Positioning System observations and revelation of seismo-ionospheric precursors of earthquakes in Total Electron Content (TEC) of the ionosphere. Global and regional relative TEC disturbances maps (%) have been plotted for 2005-2006 $M\geq6$, $D<60$ km seismic events and analyzed in order to determine general features of precursors. The obtained results agree with the recent published case-study investigations.

**Аннотация.** Перед землетрясениями наблюдаются возмущения плотности ионосферной плазмы, причиной которых является литосферно-атмосферно-ионосферное взаимодействие, в частности проникновение в ионосферу сейсмогенного электрического поля. Это поле вызывает электромагнитный $\boldsymbol{E}\times\boldsymbol{B}$ дрейф плазмы и изменение её концентрации над эпицентром и в магнитно-сопряженной области. В настоящей работе проводится анализ данных наблюдений GPS (Global Positioning System) с целью выявления сейсмо-ионосферных предвестников землетрясений в полном электронном содержании ионосферы. Для сейсмических событий 2005-2006 годов с магнитудой $M\geq6$ и глубиной залегания гипоцентра $D<60$ км были построены глобальные и региональные карты отклонений полного электронного содержания (в %) для выявления общих особенностей предвестников. Полученные в работе выводы согласуются с опубликованными ранее результатами других исследователей для конкретных сейсмических событий.

**Ключевые слова:** сейсмо-ионосферные предвестники землетрясений, полное электронное содержание ионосферы, литосферно-атмосферно-ионосферные связи, электромагнитный дрейф плазмы

## 1. Introduction

Nowadays a special attention is devoted to an investigation of earthquakes precursors observed in the ionosphere. It is known that during periods preceding strong seismic events the ionosphere can be disturbed over the epicenter region. Numerous ground-based and satellite measurements have shown the ionospheric E- and F-layer density increasing or decreasing over epicenters before seismic events (*Gokhberg et al.*, 1984; *Liperovsky et al.*, 1992; *Pulinets*, 1998; *Hayakawa et al.*, 2000; *Liu et al.*, 2000; *Ondoh*, 2000; *Depueva, Rotanova*, 2001; *Silina et al.*, 2001). The vertical electromagnetic F2-layer plasma $\boldsymbol{E}\times\boldsymbol{B}$ drift induced by a seismic electric field was proposed as the most probable physical mechanism of such anomalies formation (*Namgaladze et al.*, 2007).

Many papers mentioned electric field disturbances before earthquakes (*Kondo*, 1968; *Chmyrev et al.*, 1989; *Gousheva et al.*, 2008; *Bhattacharya et al.*, 2009; *Zhang et al.*, 2012). The problem of anomalous electric field penetration from the near-ground surface into the ionosphere has no generally accepted solution. One of the main hypotheses is ionization of the ground atmospheric gases by radon injected from the cracks of the Earth's crust. This process changes in the conductivity of the boundary layer of the atmosphere (*Pulinets et al.*, 1998; *Toutain, Baubron*, 1998). Other hypothesis is related with defect electrons also known as "positive holes" activated when rocks are subjected to ever-increasing stress (*Freund*, 2003). Then positive holes charge clouds arrive at the Earth's surface and generate a positive ground potential. Irrespective of the fact that was the source the seismic electric field the induced vertical electromagnetic drift moves plasma changing electron density over the epicenter region and in the magnetically conjugated area.

With the Global Position System distribution global ionospheric Total Electron Content (TEC) maps became available for analysis of possible earthquakes precursors. Some questions are to be investigated: whether precursors always appear before earthquakes, when precursors appear most often, whether there are local disturbances in TEC without seismic events etc.

The present paper is devoted to analysis of regional relative TEC disturbances maps (%) plotted for the earthquakes of years 2005-2006 in order to reveal earthquakes precursors and the system of their occurrence.

## 2. Methods

Analysis of TEC disturbances maps has some problems in particular with the methods of TEC data processing. There is no consensus of opinion which TEC distributions can be used as background values. Some investigations were carried out using empirically calculated background values (*Zakharenkova et al.*, 2008) or the fixed geomagnetically quiet day as a background (*Zakharenkova et al.*, 2006). Absolute TEC values were analyzed in (*Klimenko et al.*, 2011); statistical methods (computing the median and standard deviation) were applied in (*Le et al.*, 2011). 15-days medians were used in (*Zhao et al.*, 2010).

We use the Global Ionospheric Maps of TEC with resolution of 5º longitude and 2.5º latitude and temporal interval of 2 hours (*Dow et al.*, 2009). As background values for the investigated LT moment we apply moving 7-days average TEC distribution. And we analyze ratios of the current TEC pattern and the calculated undisturbed background values plotted as regional relative TEC disturbances maps.

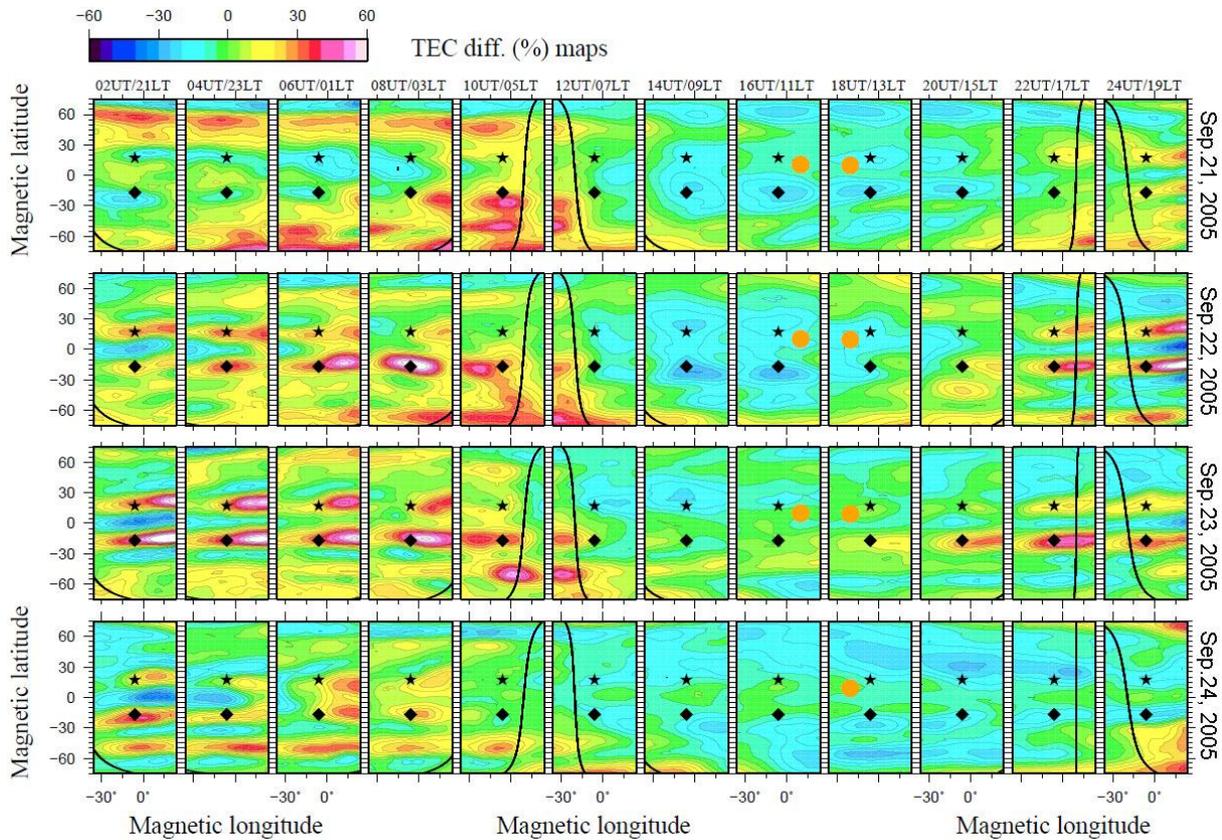

Fig. 1. TEC relative deviations (%) from the quiet background variations 4-1 days (from top to bottom) before Peru, Sep. 25, 2005, M7.5 earthquake. Co-ordinates are magnetic latitude and magnetic longitude. Star – the EQ epicenter position. Diamond – the magnetically conjugated point. Orange circle – the subsolar point. Black curve – terminator line. LT labels (above the panels) correspond to the EQ epicenter's position

Using such maps we investigate TEC disturbances before 2005-2006 earthquakes that are of M≥6 by magnitude, D<60 by hypocenter depth and <50º by epicenter geomagnetic latitude. The total number of the seismic events was 29 for year 2005 and 22 for year 2006. The sizable part of events belongs to the Oceania region and the Japan Islands region. We have analyzed the TEC ratios calculated for 1-8 days before earthquakes, because it is found out that ionospheric anomalous variations are observed 1-15 days before the strong seismic shock (*Pulinets, Boyarchuk*, 2004).

**3. Discussion**

Maps for approximately 65 % of the investigated events showed effects in TEC both at the epicenter and in the magnetically conjugated point. About 20 % of maps demonstrated positive disturbances in TEC, 12 % - negative disturbances, 14 % - both positive and negative ones. Maps for about 35 % of the considered seismic events had no local TEC precursors as disturbances both at the epicenter and in the magnetically conjugated point. Only 20 % of events have precursors which last during day-time hours. For the most part of earthquakes the TEC precursors are revealed 1-4 days before events. All maps with precursors show that the TEC disturbed regions are aligned with geomagnetic parallels of the epicenter and the magnetically conjugated point. Usually regions extend for 20º-30º in latitude and for about 30º-50º in longitude. TEC disturbances last for long periods: for 14 hours on average and 6 hours at the least. During all period of their existence the TEC precursors are located at the same regions.

Examples of the regional relative TEC disturbances maps (%) for earthquakes of 2005 are shown in Figs. 1-2. The maps demonstrate positive (Fig. 1) and negative (Fig. 2) disturbances during night hours before the events.

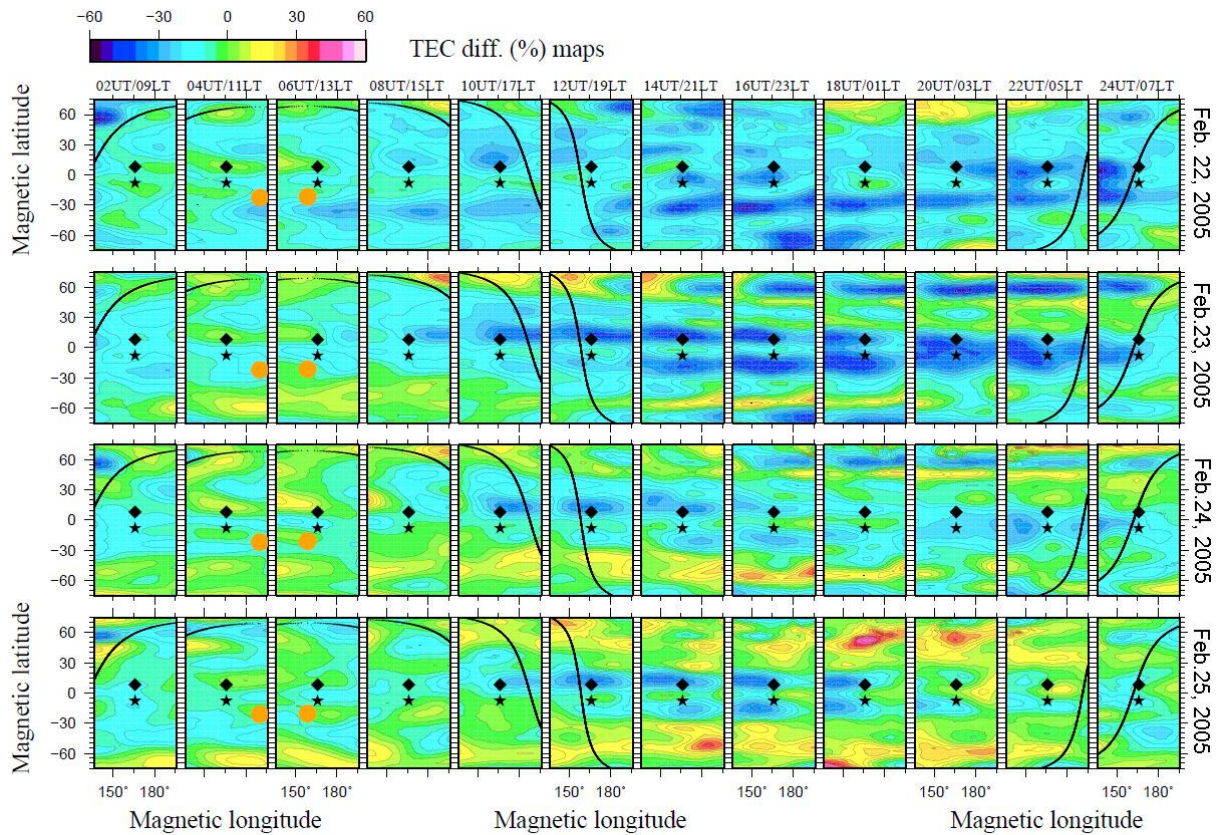

Fig. 2. TEC relative deviations (%) from the quiet background variations 4-1 days (from top to bottom) before Indonesia, Feb. 26, 2005, M6.8 earthquake. Co-ordinates are magnetic latitude and magnetic longitude. Star – the EQ epicenter position. Diamond – the magnetically conjugated point. Orange circle – the subsolar point. Black curve – terminator line. LT labels (above the panels) correspond to the EQ epicenter's position

Fig. 1 demonstrates TEC enhancements appeared 1-3 days before earthquakes. The most evident precursors belonged to the second day before the event and had the magnitude up to 60 %. The disturbances lasted from 17 LT till 05 LT. The precursor regions were aligned with the geomagnetic latitudes of the magnetically conjugated points and were characterized by the following dimensions: up to 50º in longitude and up to 20º in latitude. During daytime hours the disturbances disappeared.

In Fig. 2 disturbed by seismic impact regions had TEC lower values relative to the background. 3 days before the earthquake the disturbances were most significant. They lasted for 16 hours from 17 LT till 05 LT and had the magnitude up to -45 %. As in Fig. 1 the disturbed areas were located along parallels of the epicenter and the magnetically conjugated point. We can see shifting of the precursors regions to the geomagnetic poles because the epicenter was near the geomagnetic equator (-8º in geomagnetic latitude). So the equatorial ionization anomaly could influence on the precursor noticeably. During daytime hours the disturbances disappeared as in Fig. 1.

**4. Conclusion**

The analysis of TEC relative deviations maps has shown that most often positive TEC disturbances are revealed before seismic events. For the most part of earthquakes the TEC precursors are revealed 1-4 days before events and last for 6 hours at least and about 14 hours on average. The TEC deviation areas are aligned with the geomagnetic latitudes of the epicenter and the magnetically conjugated point. Usually such regions extend for 20º-30º in latitude and for about 30º-50º in longitude.

The investigation was performed using a significant data set (maps for more than 50 M$\geq$6, D<60 km seismic events during 2005-2006). The obtained features of the pre-earthquake TEC modifications agree with case-study investigations by *Pulinets, Boyarchuk*, 2004; *Liu et al.*, 2004; 2009; *Zakharenkova et al.*, 2006; *Pulinets et al.*, 2010; *Le et al*., 2011; *Namgaladze et al.*, 2011; *Zolotov et al.*, 2011 and support the idea on the electromagnetic mechanism of the TEC precursors generation because of the geomagnetic control and disappearance at the daytime.

**Acknowledgments.** The authors are grateful to Boris E. Prokhorov (Helmholtz Centre Potsdam, GFZ German Research Centre for Geosciences; University Potsdam, Applied Mathematics) for the help in GIM TEC maps processing.